# Conical-Shaped Titania Nanotubes for Optimized Light Management in DSSCs Reach Back-side Illumination Efficiencies > 8%

*Seulgi So,∥†Δ Arian Kriesch,∥‡Δ Ulf Peschel‡§Δ and Patrik Schmuki*†Δ*

†Department of Materials Science and Engineering, WW4-LKO, Friedrich-Alexander-University Erlangen-Nuremberg (FAU), Martensstrasse 7, 91058 Erlangen, Germany

‡Institute of Optics, Information and Photonics and Erlangen Graduate School in Advanced Optical Technologies (SAOT), Friedrich-Alexander-University Erlangen-Nuremberg (FAU), 91058 Erlangen, Germany

§Institute of Solid State Theory and Optics, Friedrich-Schiller-University Jena, 07743 Jena, Germany

ΔCluster of Excellence Engineering of Advanced Materials (EAM), Friedrich-Alexander-University Erlangen-Nuremberg (FAU), 91058 Erlangen, Germany

**Abstract**

In the present work, we introduce the anodic growth of conical shaped $TiO_2$ nanotube arrays. These titania nanocones provide a scaffold for dye-sensitized solar cell (DSSC) structures with significantly improved photon management, providing an optimized absorption profile compared with conventional cylindrical nanotube arrays. Finite difference time domain (FDTD) modelling demonstrates a drastically changed power-absorption characteristic over the tube length. When used in a back-side illumination DSSC configuration, nanocone structures can reach over 60 % higher solar cell conversion efficiency ($\eta$) than conventional tubes. The resulting $\eta \approx 8$ % represents one of the highest reported values for Grätzel type DSSCs used under back-side illumination.

Over the past decades vertically aligned $TiO_2$ nanotube layers that can be produced by a simple but optimized self-organizing electrochemical treatment of titanium in a suitable electrolyte have attracted wide scientific and technological interest.[1, 2] Due to the perpendicular alignment of these cylindrical nanotube arrays to the metal-substrate surface (back contact), a main thrust of research in this field is triggered by the anticipated beneficial use of these layers as electrodes in classic $TiO_2$ photoelectrochemical applications, such as in photocatalysis or dye-sensitized solar cells (DSSCs).[3-5] In classic Grätzel type DSSCs a photoanode is used that consists of an approximately 10 μm thick layer of $TiO_2$ nanoparticles that are coated with a monolayer of anchored dye molecules.[6-9] The dye acts as visible light absorber, with energy of the LUMO that allows excited electrons to be injected into the conduction band of $TiO_2$.[8, 10] The $TiO_2$ scaffold in this case plays only the role of a conductor for electrons to the back contact. To obtain an efficient photovoltaic conversion in DSSCs, a maximized light absorption by the dye, an optimized transport through the $TiO_2$ network and minimized recombination (mainly between oxidized dye and electrolyte) need to be established. In general, 1D structures, such as $TiO_2$ nanotubes, are considered beneficial for electron transport due to the directionality provided by their geometry.[11-13] Most straightforward is to grow $TiO_2$ nanotube directly on the Ti metal thus providing directly back-contacted electrodes. This approach allows to effectively constructing back-side illumination configurations of DSSCs. These always have a lower efficiency than front-side illumination configuration DSSCs because in this case the light passes through a Pt-coated FTO counter electrode and a longer path in the iodine electrolyte).[14] A potential shortcoming of nanotubes is their comparably low surface area. Therefore, to combine an optimum light absorption (high specific dye loading) with short connection pathways, typically hierarchical structures consisting of a $TiO_2$ 1D structure with a suitable nanoparticle decoration are used.[15-19] In classical nanoparticle-based Grätzel solar cells minimizing reflection losses, while maximizing the light absorption by designing the microstructured optical properties of the cell constituents has been achieved very successfully in experiment and theory.[20-29] Nevertheless, the efficiency of such solar cells is also determined by the shape of the depth dependent profile of photon absorption and carrier generation. A respective optimization of the light management within the absorber layer has up to now been almost completely neglected for TiO$_2$ nanotube based solar cells. Typically these layers are produced by anodization of a Ti metal sheet. Under self-organizing conditions cylindrical nanotubes arrays of TiO$_2$ with various geometries can be grown depending on the detailed anodizaiton parameters. In fact, over recent years a wide range of TiO$_2$ nanotube parameters such as length,[4, 5, 30] diameter,[30-33] and different annealing treatments[30, 34, 35] have been investigated quite extensively towards an optimized DSSC efficiency. The tubes used in these studies typically have a diameter that is in the range of 100-180 nm and a length of 5-50 μm. Some works target small diameter nanotubes (to maximize the surface area for dye coating), or use hierarchical structures where nanotubes commonly are decorated with TiO$_2$ nanoparticles. Nevertheless, virtually all studies were carried out with arrays of cylindrical tubes, as such layers are relatively easy and fast to fabricate.

Different to that, in the present article we introduce the growth of cone shaped tubes and show that with such a geometry a considerably optimized light and carrier management within a DSSC structure can be achieved in experiment and modeling. This leads to a considerable enhancement of the solar to light conversion efficiencies when these tubes are used in Grätzel type solar cells. The morphological difference between classical cylindrical tubes and cone tubes is shown in Figure 1 with high-resolution SEMs that show the tips (Figure 1 b and d) and ion-beam polished cross-section SEMs that show also the internal structure of the bases (Figure 1 c and f). Common cylindrically shaped TiO$_2$ nanotubes with a length of $L \approx 13$ μm (Figure 1d-f), were grown in 0.1 M ammonium fluoride, 5 wt % H$_2$O, 1.5M lactic acid and ethylene glycol based electrolyte.[33]

These cylinders have a very slight taper from a diameter of $Ø_b = 190 \pm 20$ nm at their base to a diameter of $Ø_t = 160 \pm 20$ nm at the top (Figure 1 d, inset). In order to grow conically shaped tubes, as shown in Figure 1a-c, we used an electrolyte based on triethyleneglycol (50 vol %), lactic acid (50 vol %) and 0.1M ammonium fluoride as described in the experimental section (see in supporting information). This electrolyte composition is the key to grow nanocone layers (Con) with a large tube diameter as shown in the SEM images of Figure 1 a-c. Here the individual tube diameter at the tube base is $Ø_b \approx 450 \pm 20$ nm tapering to a top diameter of $Ø_t \approx 250 \pm 20$ nm (Figure 1 a, inset). These conical tubes were also grown to an overall tube length of $L \approx 13$ µm. Exemplary diameters and lengths are highlighted in the high-resolution SEM zoom insets (Figure 1 b,c,e,f), while larger statistics were used to calculate the average values.

The main reason for a conical growth in this electrolyte is that the tube diameter depends on the effective anodization voltage, i.e. the applied voltage minus the resistive voltage drop in the electrolyte. In the case of the cone tubes we start with a low conductivity electrolyte that in course of the anodization process slowly self-increases its conductivity by the reaction by-products, and thus allows the average tube diameter continuously to grow, as the effective voltage continuously increases (see supporting information for more details).

Both types of tube layers were then exposed to a TiCl$_4$ treatment[8] (see supporting information) to create an optimized TiO$_2$ nanoparticle decoration, then were dye sensitized, and finally assembled to a complete DSSC (for details see supporting information). Figure 1g shows the current-to-voltage (I-V) curves for such solar cells measured under standard solar illumination AM 1.5 global (100mW/cm$^2$) conditions. The results show a remarkable difference for both tube types, yielding an efficiency of 8 % for the cone tubes as opposed to only 5 % for the cylinder tubes.

In order to clarify the reasons for this considerable efficiency enhancement in cone tubes we performed additional experiments (Figure 2) and simulations (Figure 3), particularly to pinpoint if the enhanced efficiency is connected to differences in dye loading, charge transport, or light scattering properties. Figure 2a-d show the top morphology of the tubes after the $TiCl_4$ treatment. For both tubes repeated $TiCl_4$ particle decoration was carried out until the maximum solar cell efficiency was reached. A detailed description of the process is given in Figure S2 of the supporting information. Figure 2e shows the solar cell performance for cylinder and cone tubes with and without $TiCl_4$ decoration. Table 1 provides the corresponding performance data ($J_{sc}$, $V_{oc}$, *FF*) as well as the specific dye loading determined for the different solar cells. Clearly, $TiCl_4$ decoration at both structures improves the performance drastically - namely the $J_{sc}$ is increased corresponding to a higher amount of dye loading. Nevertheless, the overall amount of dye absorbed in both nanotube geometries is virtually the same: 148 nM/cm$^2$ (Cy) and 144 nM/cm$^2$ (Con). Most importantly, please note that although the decorated tubes yield a drastic difference in efficiency $\eta$(Cy) = 5 % vs. $\eta$(Con) = 8 %, the overall optimized particle loading yields for both solar cells virtually an identical dye loading (table 1). This means that the observed differences in efficiency cannot be attributed to the potentially most straightforward reason - that is an overall different dye loading. If we additionally compare the overall reflectivity of the $TiCl_4$ treated samples we see hardly any difference (Figure 2f). A remarkable difference between Con(T) and Cy(T) is, however, that the IPCE data in Figure 2f show for the nanocones samples a clearly better performance, namely that for the long wavelength region a considerably higher IPCE is obtained. Also measurements of the electron transfer time, obtained from IMPS measurements (Figure 2h) show a significantly faster electron transport for the decorated cone tubes, while the decorated cylindrical tubes show a 3.5 times larger transport time. This improvement cannot be due to an intrinsically better electron conductivity of the cone tubes – as the non-decorated cone tubes show an even slightly inferior IPCE efficiency compared to the non-decorated cylindrical tubes.

To elucidate the origin of the clearly better performance of the cone tubes we simulated light absorption for the two geometries in the solar cells on a microscopic level, with finite difference time domain (FDTD) calculations in the extended visible wavelength range ($\lambda_0 = 400$ nm – 1 µm) (see Figure 3). All calculations were performed for an incident plane wave with the AM 1.5 global standard solar spectrum.[36] Geometry parameters for these simulations (see Figures 3a and b left) were carefully determined from experimental measurements (SEM, see Figures 1 and 2) of the $TiO_2$ cylinder and cone cells, respectively, and material parameters were chosen accordingly. Both systems were modeled as a substrate of Ti, covered with a $L = 13$ µm thick layer that consists of a rectangular close lattice of $TiO_2$ cylinders (Figure 3a), respectively cones (Figure 3b). Both types of nanotubes were covered with the experimentally optimized layer of $TiO_2$ crystals that was 25 nm thick for the cylinders and 65 nm for the cone structures - this maintains a narrow open channel in the structures, in line with experimental observation. A monolayer of dye was adsorbed into this porous matrix and the resulting sub-wavelength structured optical metamaterial was modeled according to Maxwell-Garnett-Theory (MGT)[37, 38] with a 50 % volume fraction of $TiO_2$ in anatase crystal configuration,[39, 40] and 50 % of dye. The optical properties of the dye were modeled based on experimental values from Nazeeruddin et al.[8, 9] This active layer was covered with $H_2O$ as an optical approximation of the experimental electrolyte. Light absorption in $TiO_2$ nanostructures is known to be strongly shape dependent[12, 41, 42] but has not yet been comparatively investigated for the presented nanocones and nanocylinders (more details on the procedures is given in the supporting information).

We observe in our simulations that the total absorption of solar light is almost the same for the coated nanotube cone and the nanotube cylinder system (see Figure 3d), which is well in line with the results of Figure 2f and can be understood as the total amount of dye in these systems was experimentally observed to be equal (see Figure 2, table 1). However, the spatial distribution of the light and consequently of the absorbed power is very different (Figure 3a and b, right). When the light encounters the top boundary at $z = 13$ µm, 20 – 40 % of the incident power is absorbed in the top cover layer of dye. This factor strongly depends on the exact geometry parameters, but there is no difference between cones and cylinders, and the absorption takes place at maximum distance from the bottom electrode ($z = 13$ µm).

The major difference between cones and cylinders occurs during the propagation of light through the $L = 13$ µm of the active layer formed by the array of nanotubes (Figure 3c). The material cross section of the cylinder stays nearly the same over this range, while that of the cones increases. The effect is a nearly constant differential power absorption in the cylinder system (Figure 3c blue), while the differential power absorption in the cones increases towards the bottom electrode (Figure 3c orange) adiabatically. The measured electron transfer times are shorter in case of the cones, which also indicates a more pronounced conversion close to the bottom electrode as discussed above. This is in accordance with recent observations by Zhu et al. on the reduced reflectivity of Si nanocone arrays, compared with nanowires[43] and with results by Buencuerpo et al.[44] Close to the bottom Ti electrode both absorptions reach an enhanced maximum. In other words, the main difference between the decorated cone and cylinder structure is, that maximum light absorption occurs in the cones much closer to the back electrode than in the corresponding cylinder geometry. Hence a larger amount of charge carriers is injected to the $TiO_2$ closer to the back electrode, i.e. carriers experience a short resistance path before being collected at the back contact and therefore an overall lower loss is observed.[12, 45] This is reflected in a higher IPCE efficiency of the solar cell and an average faster transport time. We therefore suggest that the increased efficiency of the cone system over that of the cylinder system is substantially caused by a shift of the absorption towards the bottom electrode for the cones.

Our hypothesis is supported by the observation of the same characteristic in the spectral absorption (Figure 2g), i.e. the total spectral absorption for incident solar radiation (normalized to the total integrated solar power) in the cylinder system and the cone system is nearly identical (closed curve). Even more importantly, the experimentally measured IPCE efficiency of the cylinder system (Figure 2g, Cy(T)) coincides with the calculated spectral absorption maximum. By analyzing the evolution of the spectral absorption along z we observe that the absorption in the cone system outperforms that in the cylinders already for an integrated absorption below $z = 6$ µm (Figure 3d, dashed curves). Furthermore we observe that the spectral absorption broadens in that range and increasingly resembles the experimental IPCE efficiency of the cones (Figure 2g, Con(T)).

Overall, the present work shows that conical $TiO_2$ nanotubes can lead to a significant enhancement in the solar light conversion efficiency in $TiO_2$ nanotube based DSSCs compared with classical cylindrical tubes. Simulations using electromagnetic finite difference time domain (FDTD) modelling show that the key reason for this improvement is the light absorption profile in cone tubes that allows for a higher amount of carriers being generated closer to the back-contact. The improved IPCE and faster average carrier transport times result in an improved efficiency of the solar cells by ≈ 60 % - leading to one of the highest values observed for back-side illuminated Grätzel type cells of ≈ 8 %.

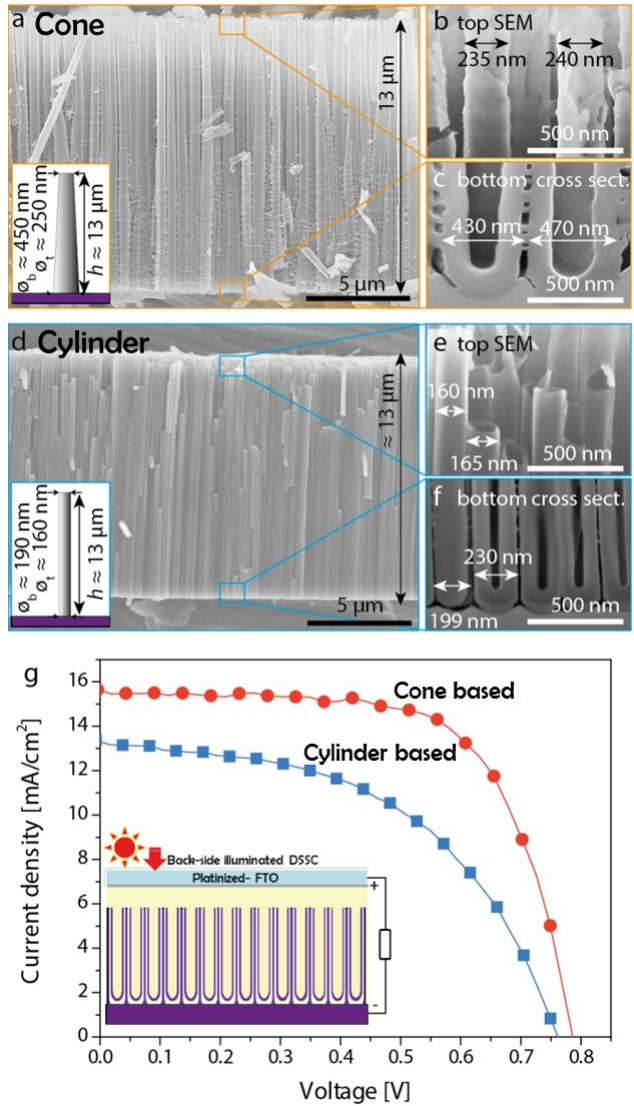

**Figure 1.** Morphological comparison of the cone (a,b,c) and cylinder (d,e,f) TiO$_2$ nanotubes with SEM. The diameter of the cones increases from $Ø_t \approx 250 \pm 20$ nm at the top (b) to $Ø_b \approx 450 \pm 20$ nm at the bottom (c) over 13µm length, while the diameter of the cylinders stays nearly constant between $Ø_t \approx 160 \pm 20$ nm and $Ø_b \approx 190 \pm 20$ nm for the same length (e, f). (g) Current-voltage (I-V) characteristics for DSSCs using both types of nanotubes.

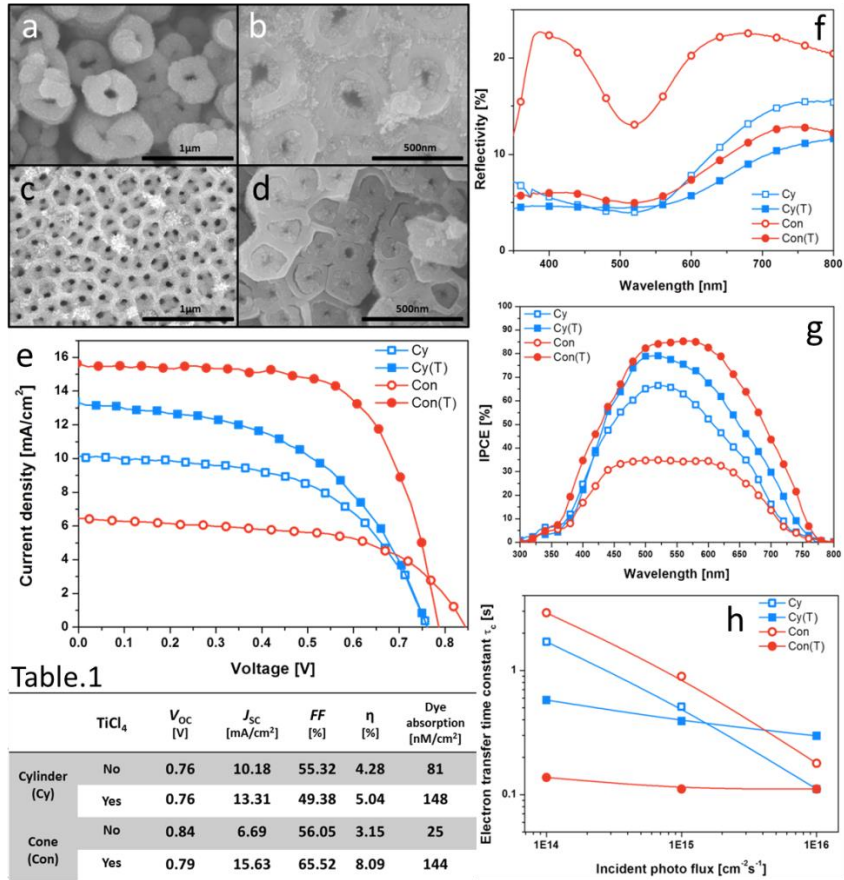

**Figure 2.** SEM images of cone nanotubes (Con) (a) top, (b) broken part near to the bottom after TiCl$_4$ decoration Con(T), and SEM images of cylinder nanotube (Cy) (c) top, (d) broken part close to the bottom after TiCl$_4$ decoration Cy(T). (e) I–V characteristics for DSSCs fabricated using Con and Cy samples with TiCl$_4$ decoration using 13 μm thickness tube layers. ($J_{SC}$ = short-circuit current, $V_{OC}$ = open-circuit voltage, $FF$= fill factor, η =efficiency) (f) Diffuse reflectivity measurement. (g) IPCE measurement. (h) Electron transfer time ($\tau_c$) constants from IMPS measurements.

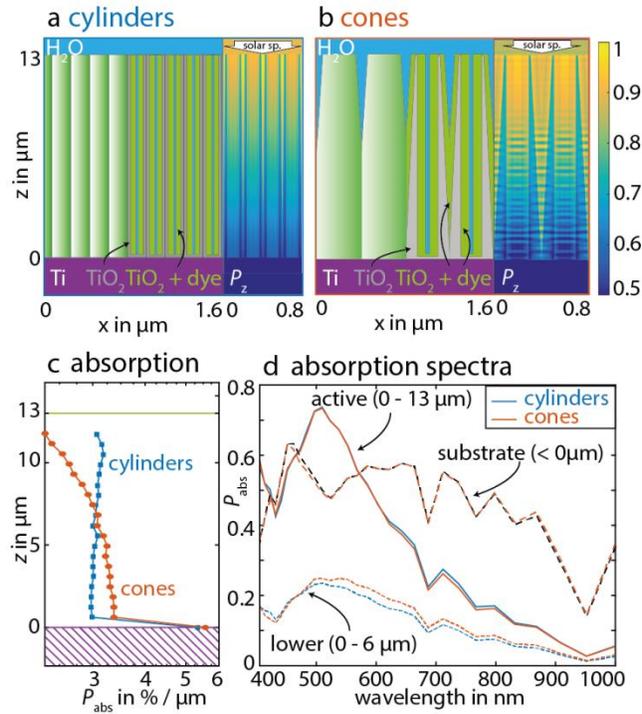

**Figure 3.** (a) Electromagnetic simulation of the $TiO_2$ cylinder system on Ti substrate, covered with a metamaterial of 50 % density $TiO_2$ crystals with dye. Power flow $P_z$ and electric field $E_x$ are indicated, as calculated with FDTD. (b) $TiO_2$ cone system, displayed like (a). (c) Differential power absorption of solar light that is launched at $z = 14$ µm and travels towards the Ti substrate. After a strong interface absorption, the absorption in the cylinder system (blue curve) remains nearly constant, while the absorption in the cone system successively increases (orange curve), before both peak at the intersection of the active layer with the bottom Ti substrate. (d) The spectral absorption profiles for the cylinders (blue) and cones (orange) resembles the experimental IPCE curves (compare Figure 2g). Within the lower 6 µm (small dashed) cone absorption outperforms cylinder absorption and the absorption spectrum broadens, in accordance with the experiment (compare Figure 2g). Residual substrate absorption is shown as large dashed.

**Supporting Information**

Electronic Supplementary Information (ESI) available: Experimental details and details on the FDTD simulations are available online. See DOI: 10.1039/c000000x/

**Author Information**

Corresponding Author

*schmuki@ww.uni-erlangen.de

Author Contributions

<sup>∥</sup>These authors contributed equally.

**Acknowledgment**

The authors acknowledge inspiring discussions with Daniel Ploss. This work was supported by the European Research council (ERC) and the Cluster of Excellence Engineering of Advanced Materials (EAM), Erlangen by the German Research Foundation (DFG) in the framework of the German excellence initiative. A.K. also acknowledges funding from the Erlangen Graduate School in Advanced Optical Technologies (SAOT) by the German Research Foundation (DFG) in the framework of the German excellence initiative.

**Abbreviations**

DSSC, dye-sensitized solar cell; FDTD, Finite difference time domain; IPCE, Incident photon-to-current efficiency; IMPS, intensity-modulated photocurrent spectroscopy; Si, silicon; FTO, fluorine doped tin oxide; $J_{SC}$, short-circuit current; $V_{OC}$, open-circuit voltage; $FF$, fill factor; η, efficiency.

**Table of contents graphic:**

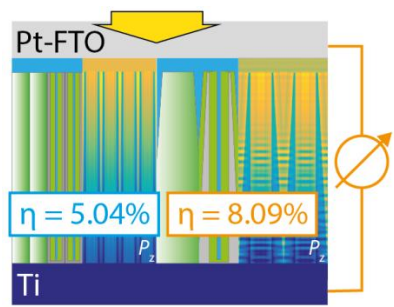

*Supporting information*

Conical-Shaped Titania Nanotubes for Optimized Light Management in DSSCs Reach Back-side Illumination Efficiencies > 8%

*Seulgi So,∥†Δ Arian Kriesch,∥‡Δ Ulf Peschel‡§Δ and Patrik Schmuki\*†Δ*

*†Department of Materials Science and Engineering, WW4-LKO, Friedrich-Alexander-University Erlangen-Nuremberg (FAU), Martensstrasse 7, 91058 Erlangen, Germany*

*‡Institute of Optics, Information and Photonics and Erlangen Graduate School in Advanced Optical Technologies (SAOT), Friedrich-Alexander-University Erlangen-Nuremberg (FAU), 91058 Erlangen, Germany*

*§Institute of Solid State Theory and Optics, Friedrich-Schiller-University Jena, 07743 Jena, Germany*

*ΔCluster of Excellence Engineering of Advanced Materials (EAM), Friedrich-Alexander-University Erlangen-Nuremberg (FAU), 91058 Erlangen, Germany*

**Experimental Section**

To grow TiO$_2$ nanotube layers we used titanium foils (0.125 mm thick, 99.6+% purity, Advent, England) that were degreased by sonication in acetone, ethanol and isopropanol, rinsed with deionized water, and then dried with a nitrogen jet. For cylinder shape nanotube layer (Cy), anodization was carried out with a high-voltage potentiostat (Jaissle IMP 88 PC) at 120 V in a two-electrode configuration with a counter electrode made of platinum gauze, using an electrolyte composition of 1.5 M lactic acid (LA, DL-Lactic acid, ~90 %, Fluka), 0.1 M ammonium fluoride (NH$_4$F) and 5 wt % deionized H$_2$O in ethylene glycol (99 vol %) held at a temperature of 60 °C (HAAKE F3 Thermostat) for 1 min.[1] The formed anodic nanotube layers from this first anodization were removed by ultra-sonication. In a second anodization, we used the same experimental conditions. For conical shaped nanotube layers (Con), anodization was carried out at 60V in two-electrode configuration using an electrolyte composed of 0.2 M ammonium fluoride (NH$_4$F), 50 vol % lactic acid (LA, dl-lactic acid, ~90 %, Fluka), and 50 vol% tri-ethyleneglycol (99 vol %) held at 50 °C with a string using a hotplate for 20h.

In order to convert the TiO$_2$ nanotubes to anatase, the samples were annealed at 450 °C in air with a heating and cooling rate of 30 °C/min during 1 h using a Rapid Thermal Annealer.

For morphological characterization, a field-emission scanning electron microscope (FE-SEM, Hitachi SEM FE 4800) was used. The thickness of the nanotubes was measured from SEM cross-sections. Further morphological and structural characterization of the TiO$_2$ nanostructures was carried out with a TEM (Philips CM30 TEM/STEM). X-ray diffraction analysis (XRD, X'pert Philips PMD with a Panalytical X'celerator detector) with graphite monochromatized CuKα radiation (Wavelength 1.54056 Å) was used for determining the crystal structure of the samples (all samples used here were fully converted to anatase).

For diffuse and specular reflectance measurements with a Lambda 950 UV/Vis/NIR spectrophotometer with a 150 mm integrating sphere (Perkin Elmer), the TiO$_2$ nanotube samples were placed at the back of the sphere.

**DSSCs:** For dye-sensitization, Ru-based dye (cis-bis (isothiocyanato) bis (2,2- bipyridyl 4,4- dicarboxylato) ruthenium(II) bis-tetrabutylammonium) (D719, Everlight, Taiwan, same as usually used "N719 dye[2]") was used. Samples were dye-sensitized by immersing them for 1 day in a 300 mM solution of the Ru-based dye in a mixture of acetonitrile and tert-butyl alcohol (volume ratio: 1:1). After dye-sensitization, the samples were rinsed with acetonitrile to remove non-chemisorbed dye. To evaluate the photovoltaic performance, the sensitized nanotubes were sandwiched together with a Pt coated fluorine-doped glass counter electrode (TCO22-15, Solaronix) using a polymer adhesive spacer (Surlyn, Dupont). Electrolyte (0.60 M BMIM-I, 0.03 M $I_2$, 0.10 M GTC in acetonitril/ valeronitril (85:15 vol.)/ SB-163, IoLiTec Inc, Germany) was introduced into the space between the sandwiched cells. Using back-side illumination, the current-voltage characteristics of the cells were measured under simulated AM 1.5 illumination provided by a solar simulator (300 W Xe with optical filter, Solarlight), applying an external bias to the cell and measuring the generated photocurrent with a Keithley model 2420 digital source meter. The active area was defined by the opening of a black shadow film-mask to be 0.2 $cm^2$. Incident photon-to-current conversion efficiency (IPCE) measurements were performed with a 150 W Xe arc lamp (LOT-Oriel Instruments) with an Oriel Cornerstone 7400 1/8 m monochromator. The light intensity was measured with an optical power meter.

For $TiCl_4$ treatments[2] we used 0.1 M aqueous solutions of $TiCl_4$ prepared under ice-cooled conditions. The $TiO_2$ nanotube layers were then treated at 70 °C for 30 min. Afterwards, the samples were washed with DI water and rinsed with ethanol to remove any excess $TiCl_4$, and finally dried in a nitrogen jet. After the treatment, $TiO_2$ nanotube samples were annealed again at 450 °C for 10 min to crystallize attached nanoparticles.

Dye desorption measurements of the dye sensitized $TiO_2$ layers were carried out by immersing the samples in 5 ml of 10 mM KOH for 30 min. The concentration of fully desorbed dye was measured spectroscopically (using a Lambda XLS UV/VIS spectrophotometer, PerkinElmer) at 520 nm and the calculated amount of dye absorption on the $TiO_2$ nanotube layer using the Beer–Lambert law.

**IMPS-IMVS:** Intensity modulated photovoltage and photocurrent spectroscopy (IMPS) measurements were carried out using modulated light (10 % modulation depth) from a high power green LED ($\lambda$ = 530 nm) and UV ($\lambda$ = 325 nm). The modulation frequency was controlled by a frequency response analyzer (FRA, Zahner IM6) and the photocurrent or photovoltage of the cell was measured using an electrochemical interface (Zahner IM6), and fed back into FRA for analysis. The light incident intensity on the cell was measured using a calibrated Si photodiode.

**Simulation:**

**Numeric simulations:** The investigated system was simulated with the finite difference time domain (FDTD) method, applying the commercial FDTD solver package FDTD solutions by Lumerical Solutions Inc. based on Maxwell's equations. The system was excited with a linearly polarized plane wave, perpendicularly incident to the sample from top (compare Figure 3a, b) with an excitation spectrum that covered $\lambda_0 = 400$ nm – 1 µm. The simulation domain was defined as a single unit cell of a rectangular lattice of the cylinders, respectively cones with geometry parameters determined from SEM analysis of the experimental samples with periodic boundary conditions in the in-plane dimensions (x, y) and perfectly matched layers (PML) constraining the simulation domain in the z dimension. After a time-to-frequency Fourier transformation of the calculated electromagnetic fields, the reflectivity and absorption in the system were carefully calculated with power integration methods and particularly full spectral absorption profiles were extracted to give the full vertical volume optical power absorption along the z domain. The spectral absorptions, as calculated for a flat spectrum, were normalized to represent the correct absorption of an incident standard solar spectrum according to the AM 1.5, ASTM G173-03(2012) atmospheric global standard.[3] Due to the $TiO_2$ band edge absorption, taking place below $\lambda_0 \approx 400$ nm, the range for spectral absorption integration that is assumed to significantly contribute to the total photovoltaic efficiency, was chosen as $\lambda_0 = 400$ nm – 1 µm. All presented spectral absorptions and spectrally integrated absorptions were globally normalized to the total integrated incident solar power within that range (Figure 3c and d).

**Material properties $TiO_2$:** The optical properties of $TiO_2$ were modeled throughout $\lambda_0 = 200$ nm – 1 µm for the anatase crystal configuration, according to experimental specifications. The experimental literature values[4, 5] in accordance with Tang et al.[6] show an absorption band edge around $\lambda_0 \approx 400$ nm, below which the optical absorption in the $TiO_2$ rapidly increases. This material was applied to the bulk $TiO_2$ that forms the cones, respectively the cylinders and for the subwavelength $TiO_2$ nanocrystals.

**Material properties dye:** The optical properties of the standard dye with ruthenium complex [Ru(dcbpyH$_2$)$_2$(NCS)$_2$] were modeled for $\lambda_0 = 200$ nm – 1 µm following experimental absorbance spectra from Nazeeruddin et al.[7] by tuning a double-Lorentzian resonance model for the spectral position and widths of the two absorption peaks at $\lambda_{0,1} = 380$ nm ± 10 nm and $\lambda_{0,2} = 515$ nm ± 10 nm.

**Material model with Maxwell-Garnett-Theory:** The composite material consisting of 50% volume fraction of TiO$_2$ nanocrystals and of 50% volume fraction of dye was modelled following Maxwell-Garnett theory (MGT).[8, 9] To match the experimental density of the dye in the composite material, the imaginary part of the permittivity was varied and the real part was recalculated to match the double Lorentzian dispersion accordingly, while the full FDTD simulations were iteratively executed and the absorption and reflection results were compared with experimental values.

Some simulation parameters, including the concentration of the dye as the active absorbing substance, were only known within certain ranges. Therefore, we conducted extensive variations in particular of the density of the dye within the Maxwell-Garnett medium to find the most reasonable configuration. A further increase of the optical density of the dye reduced the substrate absorption, but always leads to a shift of the vertical power absorption balance to the top layers. In that case the enhancement by the cones over the cylinders would probably vanish and the differential power absorption (Figure 3c) both approaches an exponential power decay characteristic.

**Figure S1**

Current density transients for cylinder (Cy) and conical (Con) shaped nanotubes in 1.5 M lactic acid, 0.1 M $NH_4F$ and 5 wt % deionized $H_2O$ in ethylene glycol electrolyte at 120 V (Cy) and in 0.2 M $NH_4F$, 50 vol % lactic acid, and 50 vol % tri-ethyleneglycol electrolyte at 60 V (Con).

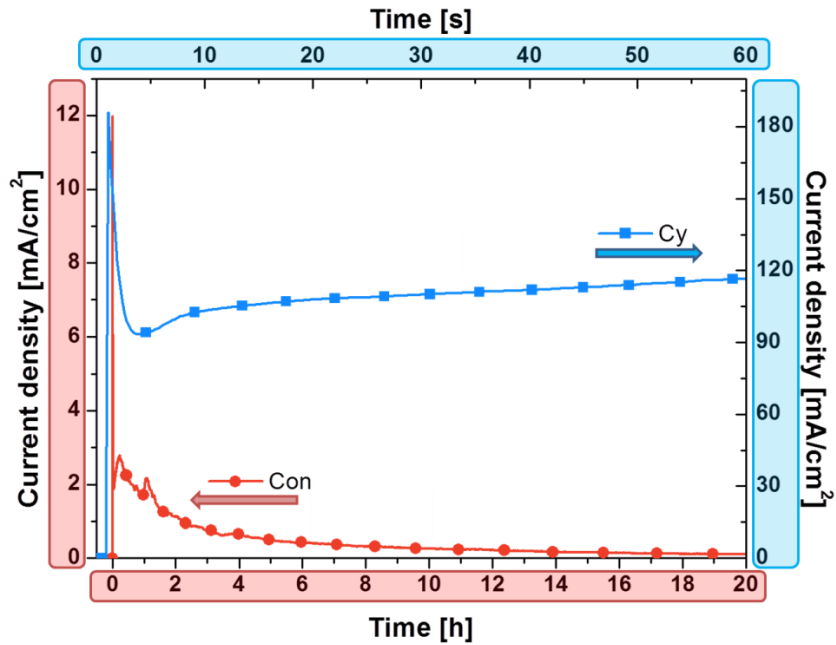

As evident from the *I-t* curve the conical shape tubes grow much slower with key difference being the electrolyte conductivity (198 µS/cm for Con-electrolyte and 440 µS/cm for Cy-electrolyte). In order to grow 13 µm cylinder tubes in the Cy-electrolyte it takes 1min (i.e. virtually no change in anodizing condition occurs). In order to grow 13 µm conical tubes it takes 20h. In this case the local changes in conductivity by reaction products lead to a permanent enhancement of the effective anodic voltage, and thus to an increase in diameter over time.

**S2-TiCl$_4$ decoration**

In order to construct efficient solar cells from such large diameter tubes (cylinder or conical), the layers were decorated with TiO$_2$ nanoparticles by a TiCl$_4$ treatment[2] to increase the overall surface area (i.e. this increases the dye absorption per unit solar cell volume). The TiCl$_4$ treatment was carried out as described in the experimental section. In each TiO$_2$ deposition cycle the outer and inner decoration thickness increases by approx. 15 nm. Crystallites typically have a size of 3 nm, and after annealing form a rigidly attached nanoparticle coating on the tube wall. For both tube types we repeated the TiCl$_4$ step until a maximum solar cell efficiency was reached (see following pages). This occurred for cylindrical tubes after 2 cycles and for conical tubes after 5 cycles (see Figure S3). Nevertheless, dye loading measurements show that in both cases almost the same amount of dye (and thus TiO$_2$ nanoparticles were deposited) to reach a maximum. Moreover, in Figure S2a, b, Con sample after 5 cycles of 0.1 M TiCl$_4$ treatment, we see that the cones are well decorated on the top and are still open. Attempts to decorate more than 7 cycles leads to partial blockage of the tubes by TiO$_2$ nanoparticles (Figure S2a). In Cy tube case only 2 cycles of TiCl$_4$ treatments could be performed without any blockage of the tubes (Figure 2c, d), and 3 cycles start to block tubes (Figure S2b). Hence, Con sample with 5 cycles of nanoparticle decoration and Cy sample with 2 cycles of nanoparticle decoration are best conditions without any blockage (and have a very similar loading of dye).

**Figure S2(a)**

Top SEM images of Con nanotube samples for different 0.1M TiCl$_4$ decoration times. (T1 = T(1time), T2=T(2times), etc.)

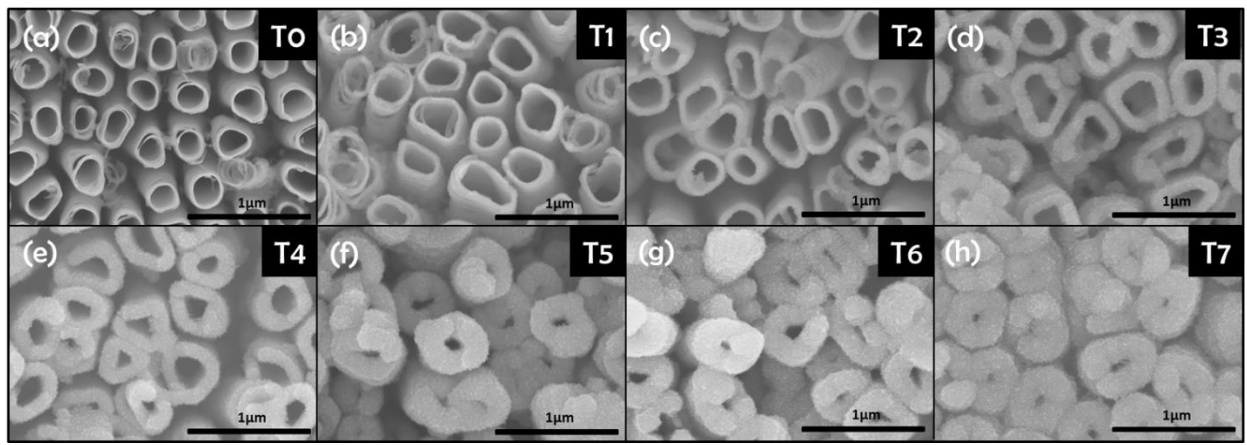

**Figure S2(b)**

Top SEM images of Cy nanotube samples for different 0.1M TiCl$_4$ decoration times. (T1 = T(1time), T2=T(2times), etc.)

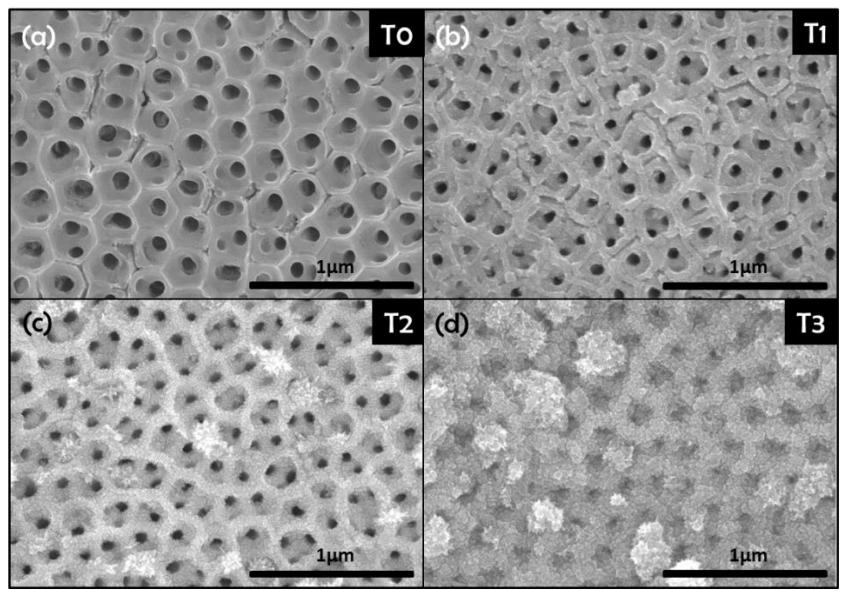

**Figure S3**

I–V characteristics for DSSCs fabricated using (a) conical and (b) cylinder nanotube samples with TiCl$_4$ decoration using 13 μm thickness tube layers. ($J_{SC}$ = short-circuit current, $V_{OC}$ = open-circuit voltage, $FF$= fill factor, η =efficiency)

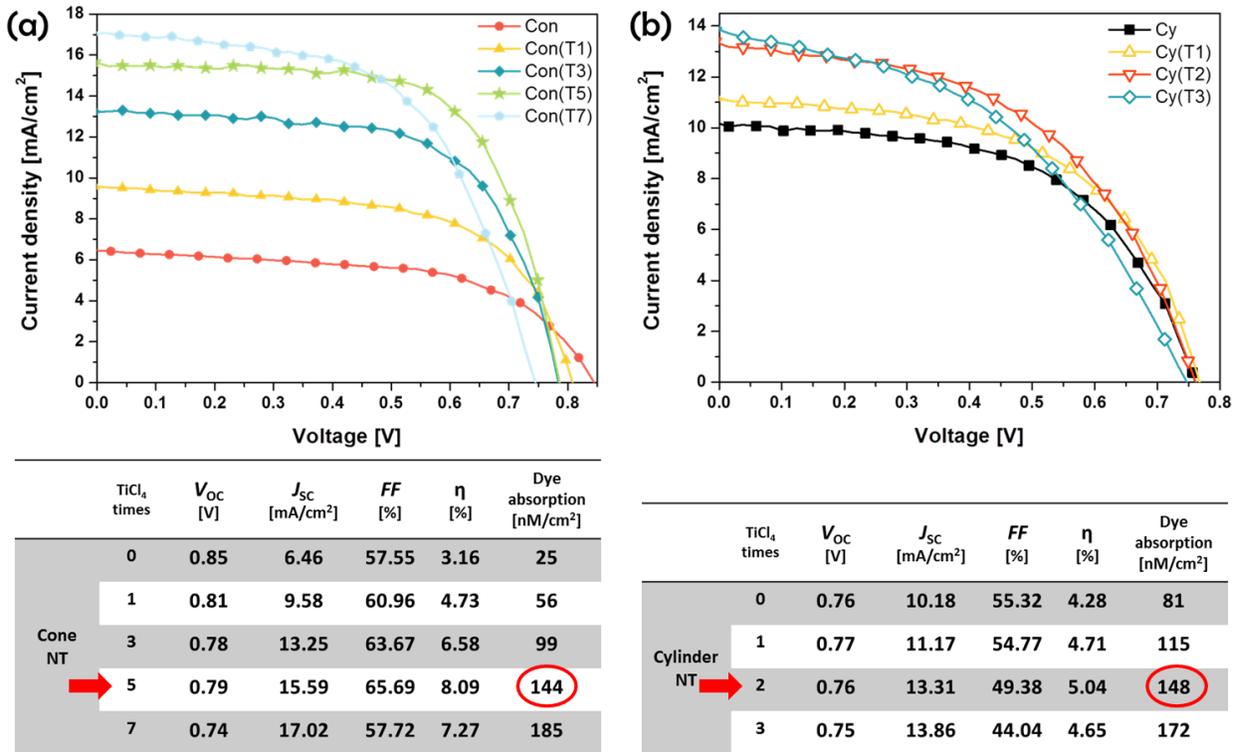

As a result we can get the highest efficiency after 5 cycles of TiCl$_4$ decoration on Con with an overall efficiency of 8.05 % with good fill factor for illumination with 1.5 solar simulators at 100 mW/cm$^2$. Higher particle loading, using 7 cycles of TiCl$_4$ decoration, still leads to an increase in $J_{SC}$ and dye adsorption, but fill factor ($FF$) is considerably decreased, which as a consequence decreases conversion results efficiency (Figure S3a).

In S3b, we can compare also TiCl$_4$ decoration properties with Cy tubes. We can observe with increasing number of TiCl$_4$ treatments, that $Jsc$ and the dye absorbance also increase - nevertheless a 2 times TiCl$_4$ treatment leads to the best solar cell efficiency of 5.04 %. After that point the results start to decrease (Figure S2b).

**Figure S4**

(a) Electron transfer time ($\tau_c$) and (b) recombination time ($\tau_r$) constants from IMPS and IMVS measurements for conical samples with $TiCl_4$ decoration.

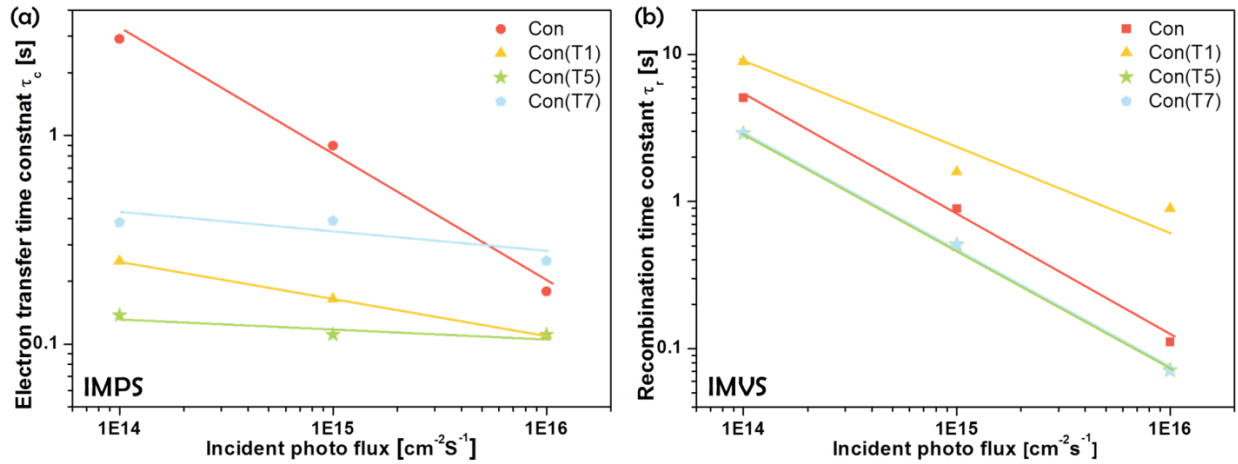

**Figure S5**

(a) Electron transfer time ($\tau_c$) and (b) recombination time ($\tau_r$) constants from IMPS and IMVS measurements for cylinder samples with $TiCl_4$ decoration.

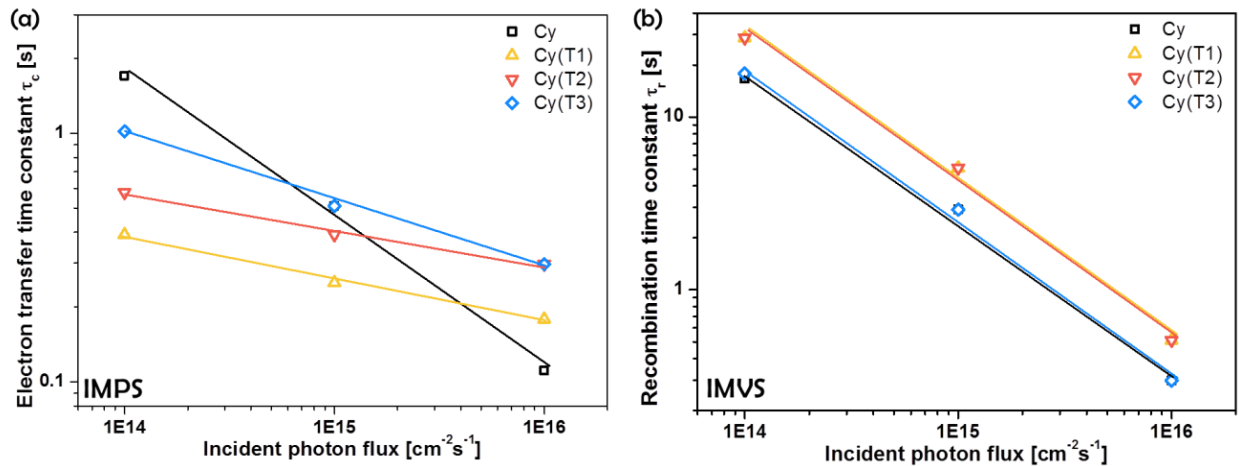

**Figure S6**

(a) Real and imaginary part of the relative permittivity of the $TiO_2$ in the spectral range of the simulation. (b) Real and imaginary part of the relative permittivity of the dye material. (c) Relative permittivity calculated from Maxwell-Garnett-Theory with 50 % density of materials as shown in (a) and (b).

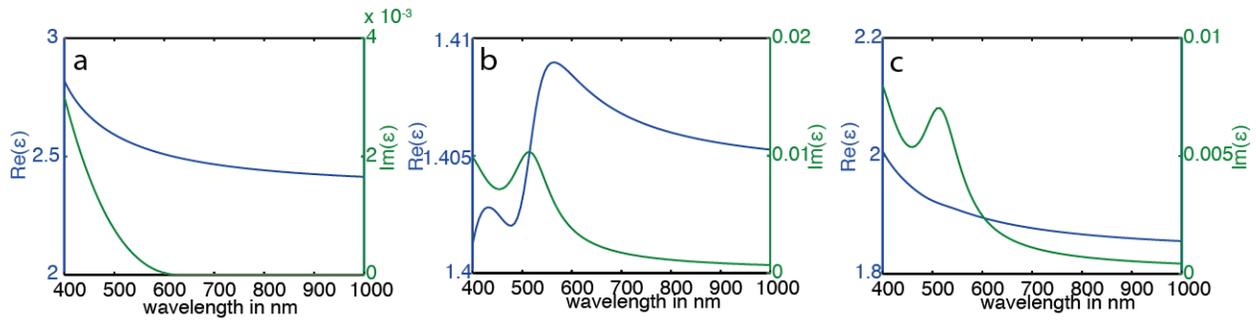

**Figure S7**

XRD spectra taken of conical and cylindrical samples with and without $TiCl_4$ treatment samples after annealed at 450 °C. (Ti=titanium, A=anatase)

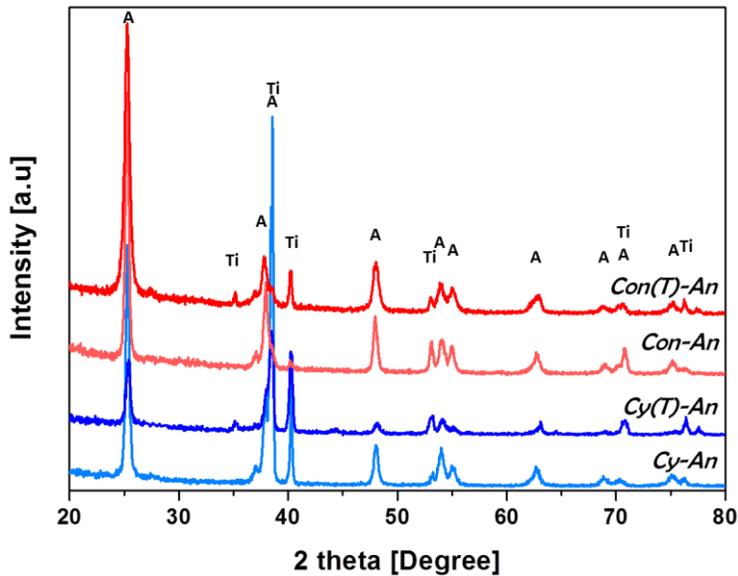